\documentclass[10pt]{wlscirep}
\title{Space-borne Bose-Einstein condensation for precision interferometry}

\author[1,+]{Dennis Becker}
\author[1,+]{Maike D. Lachmann}
\author[1,2,+]{Stephan T. Seidel}
\author[1]{Holger Ahlers}
\author[3]{Aline N. Dinkelaker}
\author[4,5]{Jens Grosse}
\author[6]{Ortwin Hellmig}
\author[4]{Hauke M\"{u}ntinga}
\author[3]{Vladimir Schkolnik}
\author[1]{Thijs Wendrich}
\author[7]{Andr\'{e} Wenzlawski}
\author[8]{Benjamin Weps}
\author[1,9]{Robin Corgier}
\author[8]{Tobias Franz}
\author[1]{Naceur Gaaloul}
\author[1]{Waldemar Herr}
\author[8]{Daniel L\"{u}dtke}
\author[1]{Manuel Popp}
\author[9]{Sirine Amri}
\author[6]{Hannes Duncker}
\author[12]{Maik Erbe}
\author[12]{Anja Kohfeldt}
\author[4]{Andr\'{e} Kubelka-Lange}
\author[4,5]{Claus Braxmaier}
\author[9]{Eric Charron}
\author[1]{Wolfgang Ertmer}
\author[3]{Markus Krutzik}
\author[4]{Claus L\"{a}mmerzahl}
\author[3]{Achim Peters}
\author[10]{Wolfgang P. Schleich}
\author[6]{Klaus Sengstock}
\author[11]{Reinhold Walser}
\author[12]{Andreas Wicht}
\author[7]{Patrick Windpassinger}
\author[1,*]{Ernst M. Rasel}
\affil[1]{Institute of Quantum Optics and QUEST-Leibniz Research School, Leibniz University Hannover, Hanover, Germany}
\affil[2]{now at OHB System AG,  We{\ss}ling/Oberpfaffenhofen, Germany}
\affil[3]{Department of Physics, Humboldt-Universit\"{a}t zu Berlin, Berlin, Germany}
\affil[4]{Center of Applied Space Technology and Microgravity (ZARM),
University of Bremen, Bremen, Germany}
\affil[5]{Department of Enabling Technologies, German Aerospace Center (DLR), Bremen, Germany}
\affil[6]{Institute of Laser-Physics, University Hamburg, Hamburg, Germany}
\affil[7]{Institute of Physics, Johannes Gutenberg University Mainz (JGU), Mainz, Germany}
\affil[8]{Simulation and Software Technology, German Aerospace Center (DLR), Brunswick, Germany}
\affil[9]{Institut des Sciences Mol\'{e}culaires d’Orsay (ISMO), CNRS, Univ. Paris-Sud, Universit\'{e} Paris-Saclay, Orsay cedex, France}
\affil[10]{Institut für Quantenphysik and Center for Integrated Quantum Science and Technology (IQ$^{ST}$), Ulm, Germany; Hagler Institute for Advanced Study at Texas A\&M University; Texas A\&M AgriLife Research; Institute for Quantum Science and Engineering (IQSE) and Department of Physics and Astronomy, Texas A\&M University, College Station, USA}
\affil[11]{Institut f\"{u}r Angewandte Physik, Technische Universit\"{a}t Darmstadt, Darmstadt, Germany}
\affil[12]{Ferdinand-Braun-Institut, Leibniz-Institut für H\"{o}chstfrequenztechnik, Berlin, Germany}
\affil[*]{rasel@iqo.uni-hannover.de}

\affil[+]{these authors contributed equally to this work}


\begin{document}
\flushbottom
\maketitle
\thispagestyle{empty}
\noindent\textbf{
Space offers virtually unlimited free-fall in gravity. 
Bose-Einstein condensation (BEC) enables ineffable low kinetic energies corresponding to pico- or even femtokelvins. 
The combination of both features makes atom interferometers with unprecedented sensitivity for inertial forces possible and opens a new era for quantum gas experiments\cite{Aguilera2014CQGSTE,CAL}. 
On January 23, 2017, we created Bose-Einstein condensates in space on the sounding rocket mission MAIUS-1 and conducted 110 experiments central to matter-wave interferometry. 
In particular, we have explored laser cooling and trapping in the presence of large accelerations as experienced during launch, and have studied the evolution, manipulation and interferometry employing Bragg scattering of BECs during the six-minute space flight. 
In this letter, we focus on the phase transition and the collective dynamics of BECs, whose impact is magnified by the extended free-fall time. 
Our experiments demonstrate a high reproducibility of the manipulation of BECs on the atom chip reflecting the exquisite control features and the robustness of our experiment.
These properties are crucial to novel protocols for creating quantum matter with designed collective excitations at the lowest kinetic energy scales close to femtokelvins\cite{Corgier}.
}
\begin{figure*}[h!]
\centering
\includegraphics[scale=1]{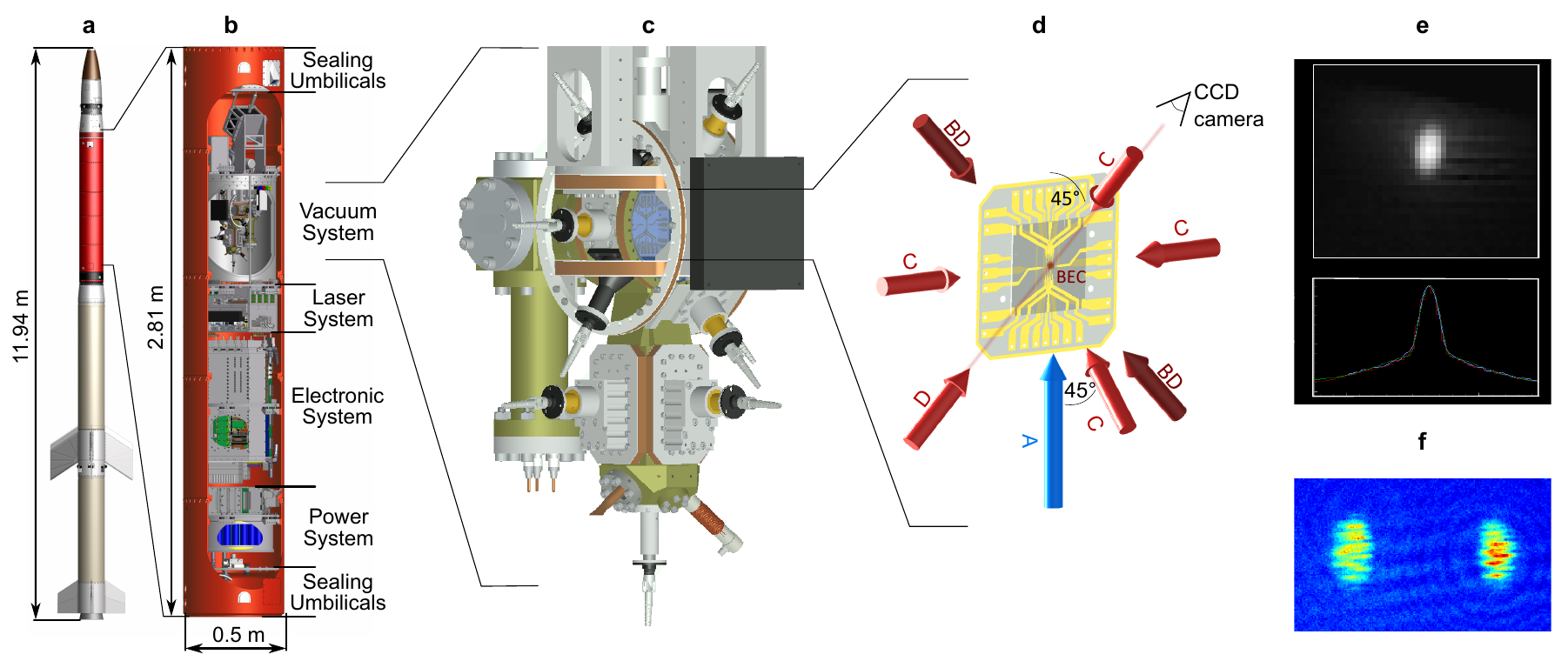}
\caption{\textbf{Setup for space-borne Bose-Einstein condensation.} \textbf{a-d},  The rocket (\textbf{a}) carried the payload (\textbf{b}) with the vacuum system (\textbf{c}) housing the atom chip (\textbf{d}) to space. On the chip, a magneto-optical trap formed by laser beams (C) is first loaded from the cold atomic beam (A). Afterwards the BEC is created in, transported by and released from the magnetic trap of the chip. Two additional light beams (BD) induce Bragg diffraction and a camera records the absorption image of a BEC using laser light (D). \textbf{e,} The picture of the first BEC in space and its one-dimensional integrated density profile were sent to ground control in low resolution. \textbf{f,} Our demonstration of Bragg scattering, apparent in the momentum distribution of the BEC, opens the path towards atom interferometry in space.}
\label{fig:setup}
\end{figure*}

Quantum systems, such as matter-waves in the presence of a gravitational field\cite{PhysRevLett.34.1472}, shine new light on our understanding of both, general relativity\cite{Misner} and quantum mechanics.
Since the sensitivity for measuring inertial forces with matter-wave interferometers is proportional to the square of the time the atoms spend in the interferometer\cite{Berman}, an extended free-fall promises an enormous enhancement in performance\cite{DimopoulosPRL2007,Aguilera2014CQGSTE}. 
In this context, Bose-Einstein condensates\cite{RevModPhys.74.875,RevModPhys.74.1131} herald a shift in paradigm because they allow us to perform interferometry over macroscopic timescales on the order of tens of seconds. 
In addition, the extreme coherence length of delta-kick collimated BECs\cite{Chu:86,PhysRevLett.110.093602,PhysRevLett.114.143004}, equivalent to temperatures as low as pico- or even femtokelvins, is mandatory to combine precision with accuracy\cite{Aguilera2014CQGSTE}. 
\begin{figure}[t!]
\centering
\includegraphics[scale=0.5]{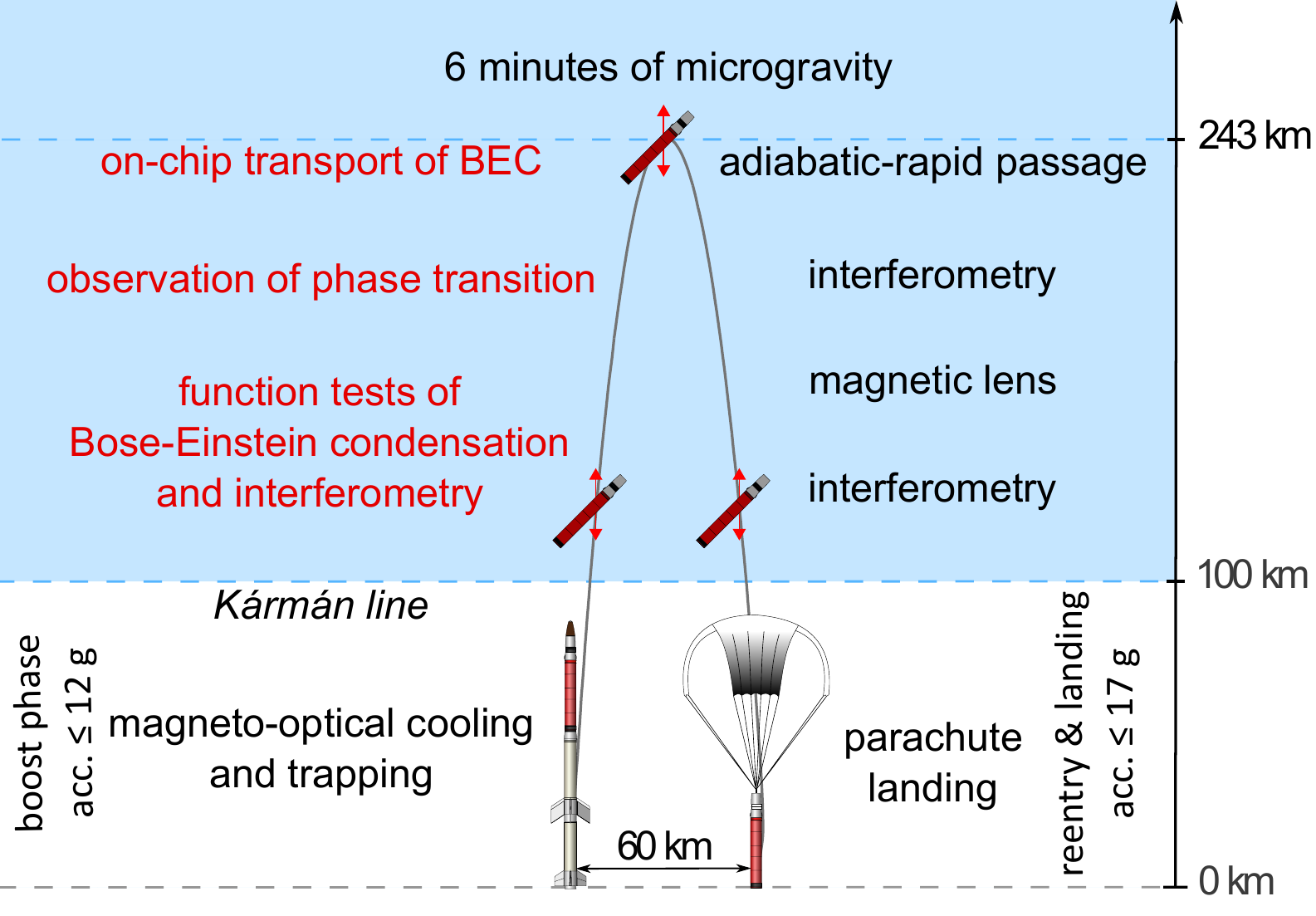}
\caption{\textbf{Schedule of the MAIUS-1 sounding rocket mission.} During its boost phase and six-minutes space flight 110 atom-optics experiments were performed.
In space, above the K\'{a}rm\'{a}n line, inertial perturbations are reduced to a few parts per million of gravity, and the spin of the rocket is suppressed to about 5\,mrad/s due to rate control. The peak forces on the payload which occur during reentry exceed gravity on ground seventeen times.
}
\label{fig:flight}
\end{figure}
\begin{figure}[h]
\centering
\includegraphics[scale=1]{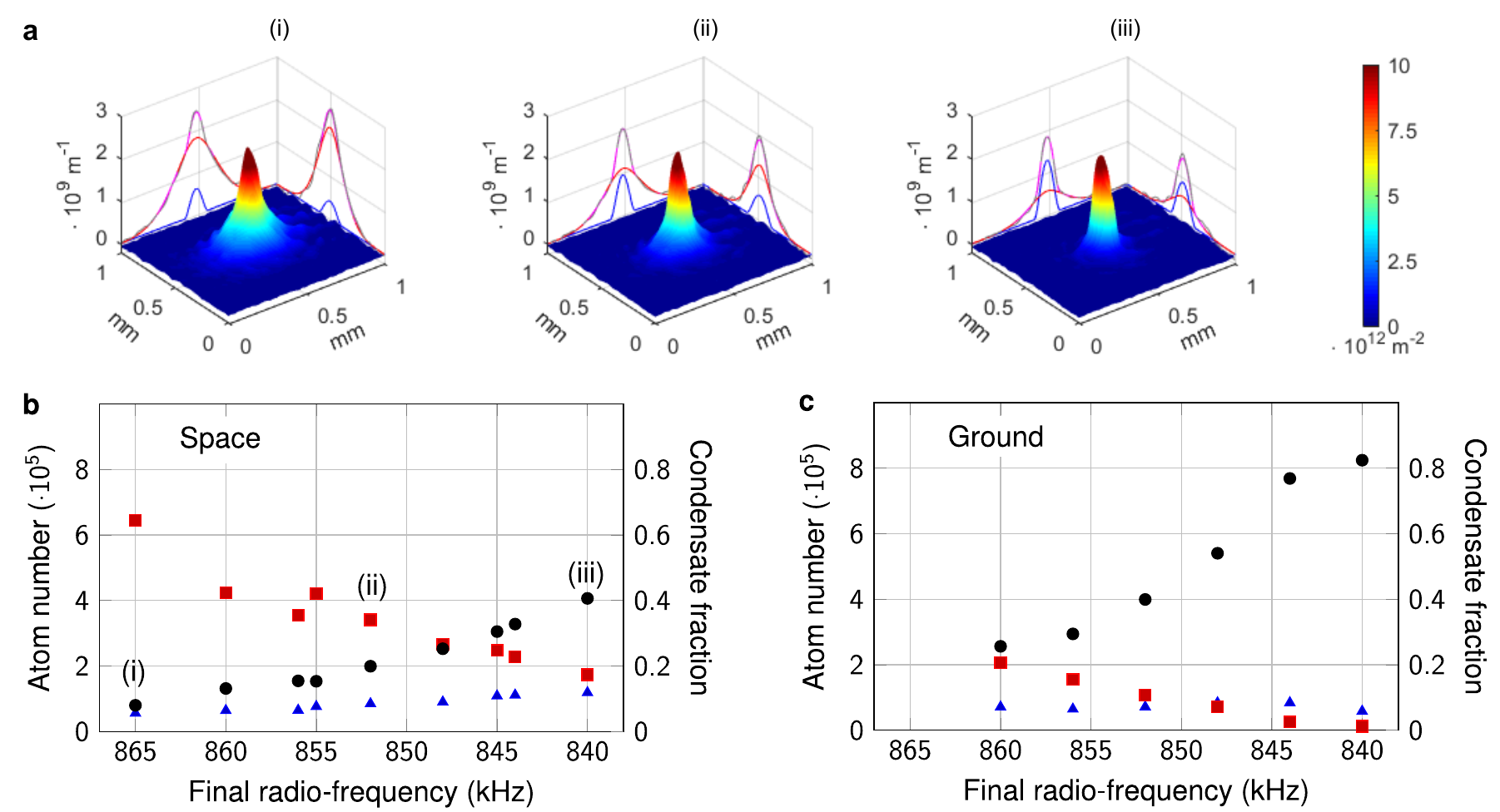}
\caption{\textbf{BEC-phase transition in space and on ground controlled by the final radio-frequency of the forced evaporation. a,} Spatial atomic densities and the corresponding line integrals depict situations in space where 8\% (i), 20\% (ii) or 41\% (iii) of the atoms are in the condensed state. \textbf{b, c,} In space (\textbf{b}) the number of magnetically trapped atoms in the thermal ensemble (red squares) is higher than on ground (\textbf{c}) resulting in more condensed atoms (blue triangles) exceeding the ground performance by 64\%. Additionally, the dependence of the condensate fraction (black dots) on the radio-frequency is different in space and on ground.
}
\label{fig:phasetransition}
\end{figure}
\begin{figure*}[p]
\centering
\includegraphics[scale=1]{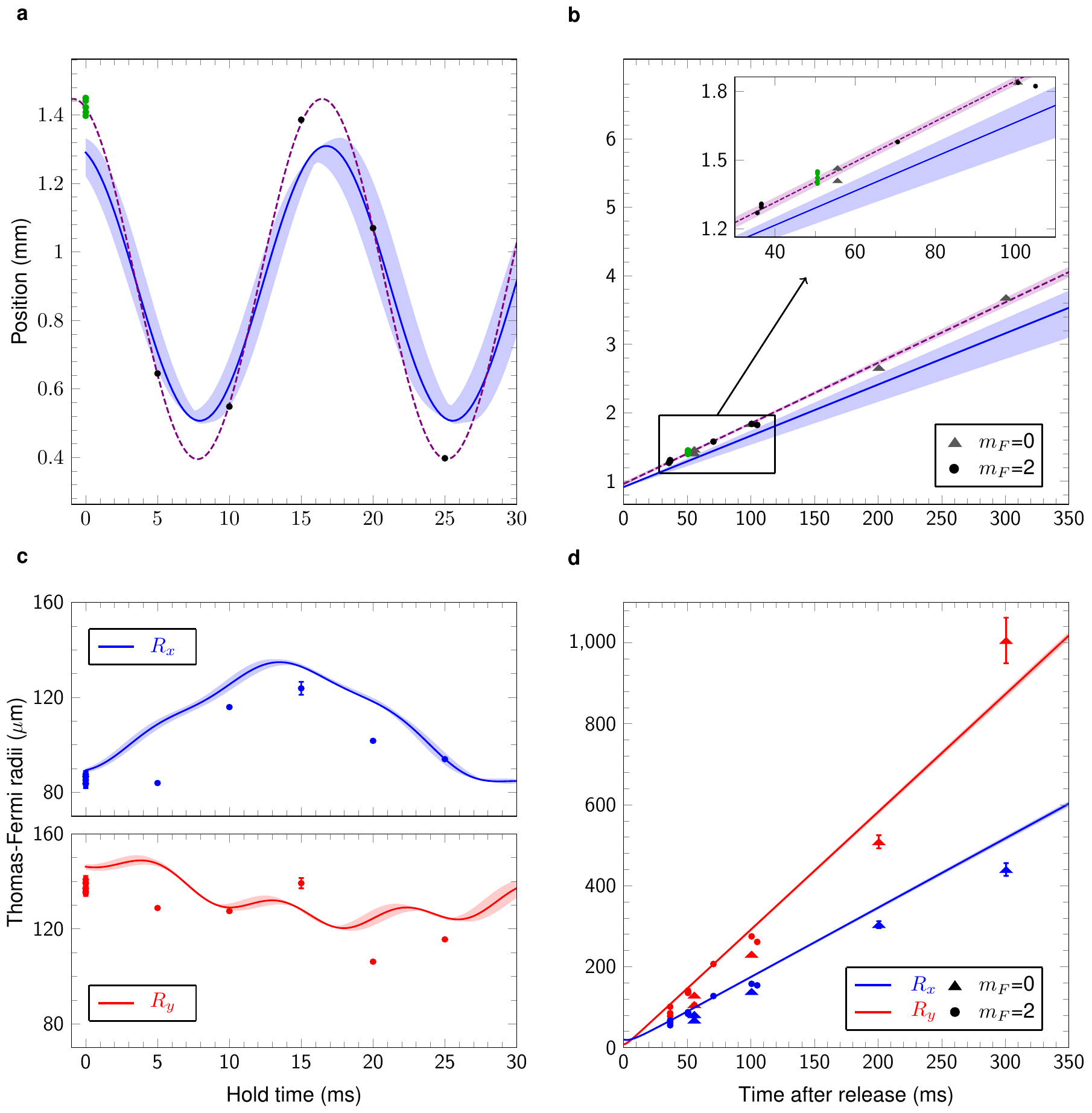}
\caption{\textbf{Excitation of the centre-of-mass motion and shape oscillations of a space-borne BEC by its transport away from the chip.} \textbf{a,} From the modulation of the distance travelled 50\,ms after release for different hold times, the transport induced oscillation is inferred by a sinusoidal fit (purple dashed line). The simulation (blue line) agrees well with the data, but underestimates the oscillation amplitude. \textbf{b,} For zero hold time which corresponds to the green data points of (\textbf{a}), the centre-of-mass motion is well fitted by the linear function shown with a 95\,\% confidence interval (purple dashed line), and almost identical for different Zeeman states of the $F=2$ manifold (grey triangles for $m_F = 0$, black and green dots for $m_F = 2$). \textbf{c,} The Thomas-Fermi radii $R_{x}$ and $R_{y}$ (blue and red dots) serve as a measure of the size and thus the shape of the BEC 50\,ms after release and, for varying hold times, they display complex oscillations in good agreement with our simulations (red and blue lines) of the BEC evolution. \textbf{d,} After 300\,ms the size of the BEC has grown for different Zeeman states (blue and red triangles and dots) to about one millimeter, in accordance with our theory (red and blue lines). Error bars and shaded areas indicate the uncertainties related to the experimental data and the theoretical model, respectively. Uncertainties in the theoretical model reflect the degree of knowledge of experimental parameters, e.g. related to the magnetic field generation by electrical circuits and currents.
}
\label{fig:4}
\end{figure*}

Despite the generation and manipulation of this state of matter being delicate, we have successfully demonstrated key methods of atom optics in microgravity on board a sounding rocket.
Our experimental apparatus\cite{doi:10.1116/1.4947583,Schkolnik2016,doi:10.1063/1.4952586} depicted in Fig. \ref{fig:setup} is equipped with a multilayer atom chip\cite{Haensel,folman2002r,fortagh2007j} and achieved an, even for terrestrial experiments, high BEC flux\cite{1367-2630-17-6-065001}.
The latter made it possible to perform a large number of experiments during the space flight, exemplified here by images of the first man-made space BEC (Fig. \ref{fig:setup}e) and Bragg scattering of a BEC (Fig.\,\ref{fig:setup}f).  

Figure \ref{fig:flight} summarises the experiments of the MAIUS-1 mission performed in space, as well as during the launch of the rocket. 
They are instrumental for NASA’s Cold Atom Laboratory\cite{CAL} (CAL) on the International Space Station (ISS) and for the NASA-DLR multi-user facility Bose-Einstein Condensate and Cold Atom Laboratory (BECCAL), which is presently in the planning phase\,\cite{BECCALWhitePaper}.

In this letter, we report on BEC experiments with Rubidium-87 atoms in space, such as the observation of the phase transition represented in Fig.\,\ref{fig:phasetransition}a by three images of spatial atomic densities of the thermal ensemble and the BEC corresponding to three characteristic final radio-frequencies of the forced evaporation. 
Figures \ref{fig:phasetransition}b,c compare the formation of BECs in space and on ground. 
Both show the respective atom numbers in the thermal ensemble (extracted with a Gaussian fit, red curve in Fig. 3a) and in the condensate (parabolic fit, blue curve in Fig. 3a), and their ratios obtained for varying temperatures, which are adjusted by terminating the evaporative cooling at different radio-frequencies.

We note that the condensate fraction observed at a given radio-frequency was lower in space than on ground, indicating a difference in magnetic field configurations. 
More remarkably, the numbers of atoms in the thermal ensemble and in the BEC in space are  64\,\% higher than those obtained on ground. 
This improvement of the BEC flux is most likely due to a more efficient loading into the magnetic trap in the absence of gravitational sag. 
To optimise the BEC flux even further the circuitry of the multilayer atom chip offers a variety of trap configurations with variable volume and depth.
However, experiments of this kind require more time than was available during our flight.

Since transporting and shaping of BECs to create compact wave packets are key to interferometry, we have investigated the transport of BECs across a distance of 0.8 mm in a sigmoidal motion of 50\,ms duration. 
By detecting the position of the BEC 50\,ms after release, for variable hold times (Fig.\,\ref{fig:4}a), the oscillation of the BEC in the release trap and its phase stability can be reconstructed.  
Out of a total of ten measurements, five tested the repeatability of the preparation for zero hold time (green dots) and five probed the oscillatory motion (black dots) for increasing hold times, all consistent with a fitted sine behaviour typical for trapped quantum gases (dashed purple line). 

Figure\,\ref{fig:4}b shows the positions of the BEC for up to 300\,ms after release and zero hold time, including data points (green dots) from Fig.\,\ref{fig:4}a for 50\,ms after release.
The release at zero hold time also implies that the oscillatory motion maximally modified the start velocity of the BEC which was as large as 8.8\,mm/s as inferred from a linear fit (dashed purple line). 
Despite the strong impact of the preparation and transport on the BEC motion, the trajectories show a remarkably small scatter in the experimental data for different Zeeman states of the $F=2$ manifold (triangles for $m_F = 0$, dots for $m_F = 2$) corresponding to a relative fluctuation in the start velocity of below 1\,\%. 
This value indicates that the various atom-chip manipulations do not affect the phase stability of the oscillation. 

The measurements are compared to a theoretical model\cite{Corgier} of the BEC dynamics after creation including its transport, release and evolution until detection. 
This model includes current-carrying wire structures of the experimental setup and solves the Gross-Pitaevskii equation in the Thomas-Fermi regime\cite{BECPethick2002}.
Our experimental results agree well with our simulation (solid blue lines), differing only by a small velocity offset in Fig 4b. 
This difference may result either from the underestimated oscillation amplitude (Fig. 4a) or from the employed model for shutting-off the release trap. 
Most other reasons are excluded by the simulation which allows us to check for influences of the circuitry of the atom chip, the Helmholtz coils and uncertainties in the current values (shaded areas in Fig. 4) on the BEC dynamics.

The transport on the atom chip causes complex shape oscillations\cite{PhysRevLett.77.2360} shown in Fig. \ref{fig:4}c, underlining again the importance of phase-stable manipulations. 
Indeed, our theoretical simulations show that the observed variations in the size of the BEC defined by the Thomas-Fermi radii actually originate from shape oscillations rather than experimental noise, as confirmed by the low phase scatter at release (Fig.\,\ref{fig:4}d) with relative fluctuations of about 2\,\%. 
Although the density of the BEC reduces by one order of magnitude during the transport, the remaining mean field energy still causes the BEC to expand up to about one millimeter in diameter after 300\,ms (Fig.\,\ref{fig:4}d), in agreement with our theory (solid blue and red lines).
The thermal background of the released atoms indicates a temperature of approx. 100\,nK and the residual expansion of the BEC corresponds to a kinetic energy equivalent to a few nanokelvins.

In conclusion, we have employed a sounding rocket payload to create BECs in space. 
From the wealth of experiments performed during this space flight, we have selected, for this letter, studies of the BEC phase transition and the collective oscillations induced by the transport away from the chip.
The demonstrated reproducibility allows the implementation of a more sophisticated transport, and to exploit shape oscillations jointly with delta-kick collimation for reducing and shaping the BEC expansion required by long interferometry times. 
We emphasise that our experiments already demonstrate the tools of atom optics required by satellite gravimetry\cite{DOUCH20181307}, quantum tests of the Equivalence principle\cite{AltschulASR2015} and gravitational wave detection based on matter-wave interferometry in space\cite{Hogan2011}.
Moreover, they pave the way to miniaturise cold-atom and photon-based quantum information concepts, and integrate them into quantum-communication satellites\cite{armengol2008quantum,ren2017ground,liao2017satellite,Yin1140,PhysRevLett.119.200501}, making the quantum internet a reality.

\bibliography{sample}
\sffamily
\footnotesize


\noindent\textbf{Acknowledgements}
This work is supported by the DLR Space Administration with funds provided by the Federal Ministry of Economics and Technology (BMWi) under grant number DLR 50WM1131-1137, 50WM0940, and 50WM1240.
W.P.S. thanks Texas A \& M University for a Faculty Fellowship at the Hagler Institute for Advanced Study at Texas A \& M University and Texas A \& M AgriLife for the support of this work. The research of the IQ$^{ST}$ is partially financed by the Ministry of Science, Research and Arts Baden-Württemberg.
N.G. acknowledges the funding from “Niedersächsisches Vorab” through the “Quantum- and Nano-Metrology (QUANOMET)” initiative within the project QT3, and WH  through the "Foundations of Physics and Metrology" project (FPM). 
R.C. is from the DAAD (Procope action and mobility scholarship) and from the IP@Leibniz program of the LU Hanover.
S.T.S is grateful for the non-monetary support of his work by DLR MORABA before, during and after the MAIUS-1 launch.
We are grateful to our former colleagues E. Kajari and M. Eckardt for the chip model code and to A. Roura and W. Zeller for their input.
We thank C. Spindeldreier and H. Blume from the IMS Hanover for FPGA software development.
We gratefully acknowledge contributions of PTB Brunswick and the LNQE Hanover for the atom chip fabrication.
We thank ESRANGE Kiruna and DLR MORABA Oberpfaffenhofen for assistance during the test and launch campaign.


\end{document}